\documentclass[twocolumn, aps, pra, showpacs]{revtex4-1}
\usepackage{eurosym}
\usepackage{graphicx}
\usepackage{dcolumn}
\usepackage{bm}
\usepackage{amsmath}
\usepackage{color}

\begin{document}

\title{Non-constant ponderomotive energy in above threshold
ionization by intense short laser pulses}

\author{R. Della Picca}
\affiliation{Centro At\'omico Bariloche, Comisi\'on Nacional de Energ\'ia
At\'omica (CONICET), Bariloche, Argentina}

\author{A. A. Gramajo}
\affiliation{Centro At\'omico Bariloche, Comisi\'on Nacional de Energ\'ia
At\'omica (CONICET), Bariloche, Argentina}

\author{C. R. Garibotti}
\affiliation{Centro At\'omico Bariloche, Comisi\'on Nacional de Energ\'ia
At\'omica (CONICET), Bariloche, Argentina}

\author{S. D. L\'opez}

\affiliation{Instituto de Astronom{\'i}a y F{\'i}sica del Espacio IAFE
(CONICET-UBA), CC 67, Suc. 28, C1428ZAA Ciudad Aut\'{o}noma de Buenos
Aires, Argentina}

\author{D. G. Arb\'o}

\affiliation{Instituto de Astronom{\'i}a y F{\'i}sica del Espacio IAFE
(CONICET-UBA), CC 67, Suc. 28, C1428ZAA Ciudad Aut\'{o}noma de Buenos
Aires, Argentina}

\date{\today}

\begin{abstract}
We analyze the contribution of the quiver kinetic energy acquired by an 
electron in an oscillating electric field to the energy balance in atomic 
ionization processes by a short laser pulse. Due to the time dependence of 
this additional kinetic energy, a temporal average is 
\textcolor{black}{assumed} to maintain a 
stationary energy conservation rule. 
\textcolor{black}{This rule is used to predict the
position of the peaks observed in the photo electron spectra (PE).}
For a flat top pulse
envelope, the mean value of the quiver energy over the whole pulse 
leads to the 
concept of ponderomotive energy $U_{p}$. 
\textcolor{black}{However for a short pulse with a fast changing field intensity a stationarity
approximation could not be precise.} 

We check these concepts by studying first the photoelectron (PE) spectrum
within the Semiclassical Model (SCM) for a multiple steps pulses. 
The SCM offers the possibility to establish a connection between emission times
and the PE spectrum in the energy domain. We show that PE substructures stem from 
ionization at different times mapping the pulse envelope. We also present the
analysis of the PE spectrum for a realistic sine-squared envelope within the
Coulomb-Volkov and \textit{ab initio} calculations solving the
time-dependent Schr\"odinger equation. We found that the
electron emission amplitudes produced at different times interfere with each other
and produce a new additional pattern, that overlap the 
above-threshold ionization (ATI) peaks.
\end{abstract}

\pacs{32.80.Rm, 32.80.Fb, 03.65.Sq}
\maketitle

\preprint{APS/123-QED} 


\section{\label{sec:level1}Introduction}


Above Threshold Ionization (ATI) processes have been extensively
studied in the context of strong laser-matter interaction 
since their first observation \cite{Agostini:79, DiMauro:95}. In these
processes the target absorbs more photons than the minimum number required
for ionization and the photo-electron (PE) spectra are characterized by
peaks at electron energy values $E_{n}$ \textcolor{black}{assumed to be} given by an energy conservation rule
\cite{Parker:96}
\begin{equation}
E_{n} + U_{p} = n\hbar \omega -I_{p},
\label{ein}
\end{equation}
where $\omega $ is the dominant laser frequency, $n$ the number of absorbed
photons, $I_{p}$ is the atomic ionization potential, and $U_{p}$ is the
denominated ponderomotive energy. This equation implicitly assumes that the
electron ejected from a stationary state with bound energy $-I_{p}$ reaches
the atomic continuum state with a defined energy determined by the action
of the laser field. However, 
\textcolor{black}{
as the Hamiltonian introduced by the laser is time dependent,  the electron final state energy is not precisely defined.} 
For a long pulse with uniform envelope, the
stationarity of the electron state can be approximately reconstituted by
assuming that 
\textcolor{black}{electron  quiver energy in the oscillating
field can be averaged over the optical cycles of the pulse resulting in the ponderomotive energy.}
 In
particular, it is well known that for an infinite flat-top pulse the 
ponderomotive energy reduces to $U_{p}=\left( F_{0}/2\omega\right) ^{2}$, 
where $F_0$ is the laser field amplitude. 
However, the concept of $U_{p}$ seems not clear for a
few-cycle intense laser pulse ~\cite{Linder:05}.
Substructures in the ATI peaks of the electron spectra due to the changing
magnitude of the ponderomotive energy along the short sine squared pulse were
found in calculations of ionization of atomic argon \cite{Wickenhauser:06}.
They considered that as the laser pulse intensity changes rapidly in a 
relatively short time period, then 
\textcolor{black}{different}
ponderomotive energy \textcolor{black}{should} 
be considered for each optical cycle.

In the present work we analyze the origin of the electron quiver energy, its
variation along the pulse and the interpretation of its mean values as
ponderomotive energies. We study how these time-dependent ponderomotive
energies are mapped on the PE energy domain. We study the dependence of the PE 
substructures 
\textcolor{black}{\cite{Wickenhauser:06}}
 on the pulse duration and shape of the envelope
stressing on the interference aspect of their formation.
Atomic ionization processes are mainly produced by tunneling near
the maximum amplitude of the electric field and, it was observed that each of these
maxima behaves as a slit, through which one electron can be emitted \cite{Linder:05}.
In analogy to the Young's experiment, we can
regard the photoelectron spectrum as the result of the interference of electron 
trajectories, giving rise to \textit{intracycle} and \textit{intercycle} interferences 
\cite{Arbo:10}. The intercycle interference stems from the contribution
\textcolor{black}{from trajectories}  in
different optical cycles and it is observed as an equally spaced set of 
\textcolor{black}{maxima} 
in the
photoelectron energy distribution. 
The intracycle interference refers to the coherent 
contribution within each optical cycle of the field and leads to 
\textcolor{black}{a equally spaced set of} peaks. 
To study the role of a variable ponderomotive energy we extend the
analytical description of the mentioned interferences to multiple-step
pulses, where each step is compose by a definite number of cycles with a
specific electric field intensity and a corresponding 
ponderomotive energy. 

This paper is organized as follows. In Sec. \ref{sec:level2} we review the concept of 
ponderomotive energy in strictly monochromatic lasers and extend it to the case of
short laser pulses. In Sec. \ref{ATI} we briefly mention the approximated methods used,
i.e., the Simple Man's Model (SMM), the Coulomb-Volkov approximation (CVA), and 
the strong field approximation (SFA), and compare their 
results with \textit{ab initio} calculation solving the time-dependent Schr\"odinger
equation (TDSE). We determine the variation of the interference pattern in the PE
spectra with the intensity of the electric field in each step. To understand
how neighboring cycles interfere we study the variation of the spectra
pattern with the temporal separation between steps. Finally, we analyze how a
time dependent ponderomotive energy can lead to the electron spectra
produced for a pulse with a realistic continuous envelope.
Atomic units are used throughout, except when otherwise stated.


\section{\label{sec:level2}Ponderomotive energy}


\subsection{Review}


In this section we review the concept of ponderomotive energy (for better
understanding of the following contents see for example chapter 2 of \cite%
{Joachain:12}). First of all we consider the classical description of an
electron in a electromagnetic (EM) field in the non-relativistic and long
wavelength (dipole) approximation. As under the dipole approximation the vector
potential can be considered spatial independent, i.e., 
$\mathbf{A}(\mathbf{r},t)\sim \mathbf{A}(\mathbf{r}_{0},t)$, 
the electric field is given by
\begin{equation}
\mathbf{F}(\mathbf{r}_{0},t)=-\frac{\mathrm{d}\mathbf{A}(\mathbf{r_0},t)}{\mathrm{d}t}
\end{equation}%
and the magnetic field 
\textcolor{black}{$\mathbf{B}= \nabla \times \mathbf{A}$ is negligible.}
Then, the electronic dynamics is governed by the Newton - Lorentz equation: 
\begin{equation}
\frac{\mathrm{d}\mathbf{v}}{\mathrm{d}t}=-\mathbf{F}(\mathbf{r}_{0},t)
\end{equation}%
and the velocity and position of the electron can be easily integrated: 
\begin{eqnarray}
\mathbf{v}(t) &=&\mathbf{A}(\mathbf{r}_{0},t)+\mathbf{v}_{d}  \label{v} \\
\mathbf{r}(t) &=&\bm{\alpha}(t)+\mathbf{v}_{d}(t-t_{0})+\mathbf{r}_{0}
\end{eqnarray}%
where the $\mathbf{r}_{0}=\mathbf{r}_{0}(t_0)$, $\mathbf{v}_{0}=\mathbf{v}_{0}(t_0)$
and $\mathbf{v}_{d}=\mathbf{v}_{0}-\mathbf{A}(\mathbf{r}_{0},t_{0})$ are 
the initial position, initial velocity, and the \emph{drift} velocity, respectively.
The displacement vector $\bm{\alpha}$ is defined by 
\begin{equation}
\bm{\alpha}(t)=\int_{t_{0}}^{t}\mathbf{A}(\mathbf{r}_{0},t^{\prime })\,%
\mathrm{d}t^{\prime }.
\end{equation}%
Therefore, according to Eq. (\ref{v}), the classical motion of an electron
in the laser field is the addition of a drift and a quiver motion
(characterized by the quiver velocity $\mathbf{A}(\mathbf{r}_{0},t)$). The
quiver kinetic energy, that is the kinetic energy that the electron acquires
by following the oscillating EM field, is given by $\mathbf{A}^{2}/2$. This
is a time dependent magnitude and the ponderomotive energy is introduced as
its cycle-averaged, or in other words, as the cycle-averaged kinetic energy
of the electron disregarding the drift velocity: 
\begin{equation}
U_{p}=\frac{1}{T}\int_{t}^{t+T}\frac{\left\vert \mathbf{A}(\mathbf{r}%
_{0},t^{\prime })\right\vert ^{2}}{2}\,\mathrm{d}t^{\prime }  \label{Up}
\end{equation}%
where $T=2\pi /\omega $ is the oscillation period of the laser
field with main frequency $\omega $. The meaning of Eq. (\ref{Up}) is clear for a
monochromatic plane wave where all cycles are identical.

Now, let us consider the quantum treatment of a free electron embedded in a
classical EM field within the dipole or long wavelength approximation. It is
know that the Volkov wave-function (or Gordon-Volkov)\cite%
{Gordon:26,Volkov:35}: 
\begin{equation}
\chi (\mathbf{r},t)=(2\pi )^{-3/2}\,\exp {\left[ \mathrm{i}\mathbf{k}\cdot 
\mathbf{r}-\frac{\mathrm{i}}{2}\int_{t_{0}}^{t}\left\vert \mathbf{k}+\mathbf{%
A}(t^{\prime })\right\vert ^{2}\mathrm{d}t^{\prime }\right] }  \label{volkov}
\end{equation}%
is a solution of the time dependent Schr\"{o}dinger equation (TDSE): 
\begin{equation}
\mathrm{i}\,\frac{\partial \big|\chi \rangle }{\partial t}=H(t)\,\big|\chi
\rangle  \label{TDSE}
\end{equation}%
where the Hamiltonian for the free electron in the EM field is $H=\left[ 
\mathbf{p}+\mathbf{A}(t)\right] ^{2}/2$. The electronic momentum $\mathbf{k} 
$ is the eigenvalue of the operator $\mathbf{p}=-\mathrm{i}\nabla $. Due to
the temporal dependence of the Hamiltonian, the energy of this state is not
defined, then the Gordon-Volkov state is not stationary and therefore the
main value of $H$ is time-dependent. Substituting Eq. (\ref{volkov}) into
Eq. (\ref{TDSE}) it is easy to finds that: 
\begin{equation}
\frac{\langle \chi \big|H(t)\big|\chi \rangle }{\langle \chi \big|\chi
\rangle }=\mathbf{k}\cdot \mathbf{A}(t)+\frac{\mathbf{k}^{2}}{2}+\frac{%
\left\vert \mathbf{A}(t)\right\vert ^{2}}{2}  \label{mainH}
\end{equation}%
A reasonable approximation for the definition of \textquotedblleft energy"
for the electron state could be achieved by averaging Eq. (\ref{mainH}) over
a cycle. This can be done in monochromatic cases or  
in flat top pulse with a main frequency and an integer number of cycles.
Recalling Eq. (\ref{Up}), we obtain: 
\begin{equation}
\frac{1}{T}\int_{0}^{T}\frac{\langle \chi \big|H(t)\big|\chi \rangle }{%
\langle \chi \big|\chi \rangle }\mathrm{d}t=\frac{\mathbf{k}^{2}}{2}+U_{p}
\label{averagemainH}
\end{equation}%
As before, we can identify the contribution of the kinetic energy coming
from the drift motion (first term of right side of the equation) and quiver
motion (last term). We note that the term 
\begin{equation}
\beta (t)=\frac{1}{2}\int_{t_{0}}^{t}\left\vert \mathbf{A}(t^{\prime
})\right\vert ^{2}\,\mathrm{d}t^{\prime }  \label{beta}
\end{equation}%
in the phase of the Volkov state Eq. (\ref{volkov}) is the responsible of
the quiver kinetic energy contribution in Eq. (\ref{mainH}) and (\ref%
{averagemainH}). In fact, the ponderomotive term [Eq. (\ref{beta})]
is omitted via the velocity gauge transformation (see \cite{Joachain:12}) 
and thus, the velocity-Volkov state $\chi^{V}=\chi \,\exp {[\mathrm{i}\beta (t)]}$
and the new velocity-Hamiltonian
$H^{V}= \mathbf{p}^{2}/2+\mathbf{p}\cdot \mathbf{A}(t)$
do not reproduce the term $U_{p}$ in Eq. (\ref{averagemainH}). Otherwise,
under length gauge transformation, the state 
\begin{equation}
\chi ^{L}(\mathbf{r},t)=\chi (\mathbf{r},t)\,\exp {[\mathrm{i}\mathbf{A}%
(t)\cdot \mathbf{r}]}  \label{volkov_L}
\end{equation}%
and the Hamiltonian $H^{L}=\mathbf{p}^{2}/2+\mathbf{r}\cdot \mathbf{F}(t)$
verify also Eq. (\ref{averagemainH}).

A few-cycle laser pulse not only cannot be considered as monochromatic since the
finite duration of the pulse generates a spread in the frequency domain, but also
generally does not have identical optical cycles due to the time dependent envelope.
Therefore the concepts introduced in this section should be revised.


\subsection{Envelope dependence}


For a short pulse, the time-dependence of the envelope must be considered,
thus the definition of the ponderomotive energy  must be revised.
When the characteristic time of the envelope variation is much higher than 
the laser period (slow variation), we can still evaluate the time average energy
over each optical cycle. In particular, we investigate laser pulses with a
main frequency $\omega $, in spite of its finite duration. Thereby, it is
possible to define an oscillation period as $T=2\pi /\omega $, and then the $%
m$th cycle is contained at the interval time between $(m-1)T$ and $mT$, with 
$m=1,...,N$ and $N$ is the total number of cycles. Then, the cycle-average
of the energy in Eq. (\ref{mainH}) depends on $m$ as
\begin{equation}
\frac{1}{T}\int_{(m-1)T}^{mT}\frac{\langle \chi \big|H(t)\big|\chi \rangle }{%
\langle \chi \big|\chi \rangle }\mathrm{d}t=\frac{\mathbf{k}\cdot \Delta %
\bm{\alpha}(m)}{T}+\frac{\mathbf{k}^{2}}{2}+U_{p}(m),
\label{averagemainHm}
\end{equation}%
where we have introduced an optical-cycle dependent ponderomotive energy as 
\begin{equation}
U_{p}(m)=\frac{1}{T}\int_{(m-1)T}^{mT}\frac{\left\vert \mathbf{A}(t^{\prime
})\right\vert ^{2}}{2}\,\mathrm{d}t^{\prime }  \label{Upm}
\end{equation}%
where $m$ indicates the $m$th optical cycle, and 
\begin{equation}
\Delta \bm{\alpha}(m)=\bm{\alpha}(mT)-\bm{\alpha}((m-1)T)=\int_{(m-1)T}^{mT}%
\mathbf{A}(t^{\prime })\,\mathrm{d}t^{\prime }\,.  \notag
\end{equation}%
When the envelope is constant or varies slowly, the term $\Delta \bm{\alpha}%
(m)$ could be considered negligible since the vector potential has the
positive contribution practically equal to the negative one during an
oscillation cycle. On the other hand, the temporal integration of $\mathbf{A}
$ over all duration of the pulse should be zero 
due to the finite size of the laser oscillator cavity
 \cite{Joachain:12}, then $\sum_{m}\Delta \bm{\alpha}(m)=0$.

\begin{figure}[tbp]
\includegraphics[width=0.5\textwidth]{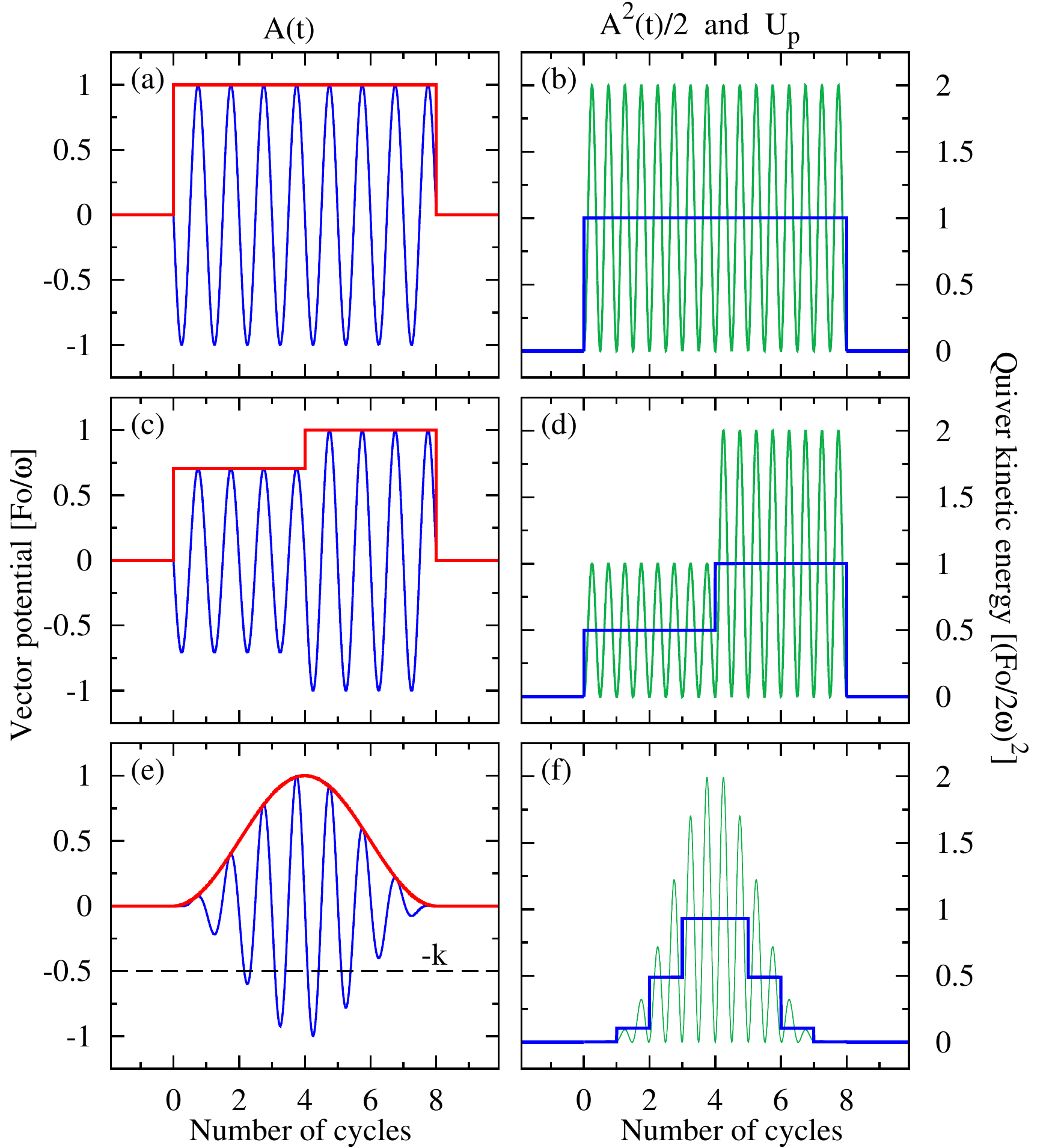}
\caption{(Color online) Vector potential and quiver kinetic energy as
function of the number of cycles $m=\protect\omega t/2\protect\pi $, for
three envelopes: flat top, 1 step and sine squared envelope (top, middle and
lower panels respectively). The laser pulse duration is equivalent to 8
cycles. Here $F_{1}=F_{0}/\protect\sqrt{2}$, $F_2=F_0$ and $t_{1}=t_2=%
\protect\tau /2$. Left column: Vector potential amplitude $A(t)$ in thin
(blue) line and envelope of the laser field thick (red) line in units of $F_0/\omega$. Right column:
quiver kinetic energy $A^{2}(t)/2$ in thin (green) line and cycle averaged
quiver kinetic energy $U_{p}(m)$ in thick (blue) line in units of $(F_0/2\omega)^2$.}
\label{Vs_env}
\end{figure}

In order to illustrate the dependence of the ponderomotive energy with the
envelope, let us consider a linearly polarized laser pulse modeled by the electric field 
\begin{equation}
\mathbf{F}_{i}(t)=f_{i}(t)\,\cos {(\omega t)}\,\mathbf{\hat{\varepsilon}\,},
\label{field}
\end{equation}%
where we choose linear polarization along the direction $\hat{\varepsilon}$
and the total pulse duration $\tau =NT$ is such that the field oscillates $N$
cycles. We consider three different types of envelope $f_{i}(t)$ defined as 
\begin{eqnarray}
f_{1}(t) &=&F_{0}  \label{f1} \\
f_{2}(t) &=&%
\begin{cases}
F_{1}, & \text{if }0<t<t_{1} \\ 
0, & \text{if }t_{1}<t<t_{2} \\ 
F_{2}, & \text{if }t_{2}<t<\tau%
\end{cases}
\label{f2} \\
f_{3}(t) &=&F_{0}\,\sin ^{2}(\pi t/\tau ),  \label{f3}
\end{eqnarray}%
and zero outside the time interval $[0,\tau ]$.
Figure \ref{Vs_env} presents the envelope, potential vector, quiver and ponderomotive
energy for these three laser pulses for $N=8$.
In the first case (Fig. \ref{Vs_env}(a) and (b)) corresponding to the flat top
envelope, all cycles are identical to each other, the mean
ponderomotive energy is equal to $(F_{0}/2\omega )^{2}$ along the pulse
in a plane wave laser. Instead, the two step [$t_1 = t_2$ in Eq. (\ref{f2})]
and sine squared [Eq. (\ref{f3})] envelopes show clearly the
dependence of $U_{p}$ on the 
$m$th cycle where the average is performed [see Fig. \ref{Vs_env}(d) and (f)].
However, to the best of our knowledge, for few-cycle laser pulses
it is common to define in the literature a unique value of the ponderomotive energy as
$U_p^0=(F_0/2 \omega)^2$ with $F_0$ the maximal value of the envelope despite
its strong envelope time variation.

Another important issue is about $\Delta \bm{\alpha}(m)$. As we have noted
before, in the flat top [Eq. (\ref{f1})] and step [Eq. (\ref{f2})] cases,
it is clear that $\Delta \bm{\alpha}(m)=0$
since the envelope remains constant during each cycle, whereas for
the $f_{3}$ envelope it can be shown that $\Delta \bm{\alpha}(m)$
decreases as $N$ increases and it becomes negligible at the
middle of the pulse ($m \sim \frac{N}{2}$) where the laser intensity (and then
the probability of electronic emission in ionization problems) is maximum.


\section{Above threshold ionization}
\label{ATI}

We consider the problem of above threshold ionization of one-electron
atom due to the interaction with a classical EM field, in particular, short
laser pulses. It is known that ATI peaks produced by long pulses in the PE
spectra are located in the positions given by Eq. (\ref{ein}). 
Its validity requires 
\textcolor{black}{ an initial and final state energies well defined.}
However,
when a laser pulse impinges on an atom, the electron leaves its stationary
state and its energy level changes over time and the simple relationship
given by Eq. (\ref{ein}) is not exact for this time-dependent process.
A reasonable approximation can be
achieved by averaging the 
\textcolor{black}{main value of $H$ [Eq. (\ref{mainH})]} 
over a pulse optical cycle. As we
have already explained, this is exact for a monochromatic pulse with flat
envelope and the pulse has an integer number of cycles.
When the pulse
envelope varies in time it is not clear the concept of energy conservation
Eq. (\ref{ein}) and the ensuing exact position of ATI peaks. 
For this
reason, we will analyze the electron emission spectra and the validity of
Eq. (\ref{ein}) in cases where the envelope results on a varying
ponderomotive energy. We examine ionization of a hydrogen atom from the $1s$
state by short laser pulses with frequency and amplitude parameters such
that the value $(F_{0}/2\omega )^{2}$, as indicator of the ponderomotive
energy, is comparable to the separation of ATI peaks, \textit{i.e.} $\omega $
and, therefore, Eq. (\ref{ein}) results sensitive to variations of the
ponderomotive energy.

To investigate the PE spectra, especially within approaches based on time
dependent quantum theory, the Volkov sate given by Eq. (\ref{volkov}) has
been widely used to describe the emitted electron immersed in the laser
field. For example the Strong Field Approximation (SFA) \cite%
{Keldysh:64,Faisal:73, Reiss:80, Milosevic:06}, the Coulomb-Volkov
Approximation (CVA) \cite{Duchateau:02,Rodriguez:04,Arbo:08,
Gravielle:12,DellaPicca:13} and the semiclassical application of the Simple
Man's Model (SCM) 
\cite{Arbo:10R,Arbo:10,Arbo:14} employ the non-stationary Volkov state as
final wavefunction. 
The use of length gauge
for the perturbation potential, \textit{i.e.} $\mathbf{r}\cdot \mathbf{F}(t)$, and the Volkov state Eq. (\ref{volkov_L}) also
guarantee the presence of $U_{p}$ in the \textquotedblleft final
energy\textquotedblright, as we have mentioned before.

Within the time-dependent distorted wave theory, the transition amplitude in
the prior form and length gauge is expressed as 
\begin{equation}
T_{if}=-i\int_{0}^{\tau }\big\langle\chi _{f}^{-}(\mathbf{r},t)|\mathbf{r}%
\cdot \mathbf{F}(t)|\Phi _{i}(\mathbf{r},t)\big\rangle\,\mathrm{d}t
\label{T}
\end{equation}%
where $\Phi _{i}(\mathbf{r},t)=\varphi (\mathbf{r})\,\mathrm{e}^{\mathrm{i}%
I_{p}t}$ is the $1s$ initial state of the hydrogen atom and $\chi _{f}^{-}(%
\mathbf{r},t)$ is the length Volkov state Eq. (\ref{volkov_L}) in the SFA.
In the CVA, $\chi _{f}^{-}$ includes the coulomb distortion as an additional
factor: 
\begin{equation}
\chi _{f}^{-}(\mathbf{r},t)=\chi ^{L}(\mathbf{r},t)\,D_{C}^{-}(\mathbf{k},%
\mathbf{r})
\end{equation}%
where $D_{C}^{-}(\mathbf{k},\mathbf{r})=N(\nu )\,{_{1}F_{1}}\left( \mathrm{i}%
\nu ;1;-\mathrm{i}(kr+\mathbf{k}\cdot \mathbf{r})\right) $, the
normalization factor is defined as $N(\nu )=\Gamma (1-\mathrm{i}\nu )\mathrm{%
e}^{-\pi \nu /2}$ and $\nu =-1/k$ is the Sommerfeld parameter, and ${%
_{1}F_{1}}$ is the hypergeometric function.

The photoelectron momentum distributions is obtained from the transition
matrix magnitude as 
\begin{equation}
\frac{\mathrm{d}P}{\mathrm{d}\mathbf{k}}=|T_{if}|^{2}  \label{P}
\end{equation}%
and the PE spectrum, that is the differential ionization probability of
emitting electron with energy $E=k^{2}/2$, is 
\begin{equation}
\frac{\mathrm{d}P}{\mathrm{d}E}=\int k\,|T_{if}|^{2}\mathrm{d}\Omega _{k}.
\label{PE}
\end{equation}

The semiclassical model (SCM) described in
previous works \cite{Arbo:10R,Arbo:10,Arbo:14}, utilizes the saddle point
approximation to replace the time integration in Eq. (\ref{T}) by a sum over
ionization or release times $t_{r}$ where the oscillating phase 
\begin{equation}
S(t)=\int_{0}^{t}\left[ \frac{(\mathbf{k}+\mathbf{A}(t^{\prime }))^{2}}{2}%
+I_{p}\right] \mathrm{d}t^{\prime }
\label{action}
\end{equation}%
is stationary and has null derivative. Here, $S$ represents the modified
Volkov action in which the energy of the initial state is included.
The SCM approximates the complex release times $t_{r}$ with the real solutions of 
\begin{equation}
\mathbf{k}+\mathbf{A}(t_{r})=0\,,  \label{tr}
\end{equation}%
assuming that the electron is emitted from the atom into the continuum at
the release time $t_{r}$ with zero initial velocity. 


\subsection{Flat-top pulse}


The coherent superposition of classical trajectories associated with release times $%
t_{r}^{i}$ $(i=1,2,\dots )$ gives rise to semiclassical interferences,
provided they satisfy the condition given by Eq. (\ref{tr}) for reaching the
same final momentum $\mathbf{k}$. For a flat-top pulse the transition matrix can be written
as
\begin{equation}
T_{if}=G(\mathbf{k})\sum_{i=1}^{M}\exp {\left[ \mathrm{i}S(t_{r}^{i})\right] 
}\,,  \label{T_sum_i}
\end{equation}%
where $M$ is the number of classical trajectories reaching a given final
momentum $\mathbf{k}$ and $G(\mathbf{k})$ is the ionization amplitude that
is independent of the emission times.

For an $N$-cycle flat top pulse (envelope defined in Eq. (\ref{f1}) and Fig. 1 a)
there are two time solutions of Eq. (\ref{tr}) for each cycle $m$: $t_{1}^{m}$
and $t_{2}^{m}$. Let us introduce the difference and the average of the action
evaluated at these times: 
\begin{eqnarray}
\Delta S &=&S(t_{2}^{m})-S(t_{1}^{m})  \label{deltaS} \\
\overline{S}_{m} &=&\frac{S(t_{2}^{m})+S(t_{1}^{m})}{2}=S_{0}+m\bar{S}
\end{eqnarray}%
where $\Delta S$ is the accumulated action, independent of the cycle $m$ and $%
\overline{S}_{m}$ is the average action in the $m$-cycle that depends linearly
with $m$, where
$\bar{S} =\frac{2\pi }{\omega }\left( \frac{k^{2}}{2}+U_{p}+I_{p}\right)$ and
$S_{0} =-\bar{S}/2-kF_{0}/\omega^{2}$.
We can derive an analytical expression for the transition amplitude
as \cite{Arbo:10R,Arbo:10}
\begin{equation}
T_{if}=2\,G(\mathbf{k})\,\cos {\left( \frac{\Delta {S}}{2}\right) }\,\frac{%
\sin (N\bar{S}/2)}{\sin ({\bar{S}/2})}\,\mathrm{e}^{\mathrm{i}(S_{0}+\frac{%
N+1}{2}\bar{S})}\,,  \label{TSM}
\end{equation}%

When the square modulus of Eq. (\ref{TSM}) is considered, the factor $\cos
^{2}(\Delta {S}/2)$ represents the form factor (or structure factor)
accounting for intracycle interference, since it arises from the pairs of
classical trajectories born at different times separated by $\Delta t =t_{2}^{m}-t_{1}^{m}$
between them. On the other hand, the factor 
$\left[ \sin (N\bar{S}/2)/\sin ({\bar{S}/2})\right] ^{2}$ is the responsible for
the intercycle interference, resulting
in the well-known ATI peaks. In fact, when $N\rightarrow \infty $ this
factor become a sequence of delta functions situated at $E_n$ values given by
Eq. (\ref{ein}).


\subsection{\label{SMM}Semiclassical model for interpulse interference}


For a more clear and comprehensive analysis of how the shape of the pulse envelope
 affects the PE spectra we assume an $N$-cycle laser pulse modeled by
the simple time-dependent envelope $f_{2}(t)$, defined by two flat-top
pulses with different amplitudes (as in Fig. \ref{Vs_env} c-d) 
and separated by a time interval $\Delta t=t_{2}-t_{1}$. 
This last condition allows us to see how the interference pattern
depends on the temporal separation between pulses. 
We assume that the time gap of $\Delta t$ corresponds to $M=\Delta t/T$ cycles,
with $T=2\pi /\omega$. The first pulse corresponds to a $N_{1}$-cycle flat-top pulse
with a field strength $F_{1}$ and the second to a $N_{2}$-cycle flat-top pulse
with a field strength $F_{2}$, where $N_{1}=t_{1}/T$, $N_{2}=(\tau -t_{2})/T$
and $N=N_{1}+N_{2}+M$.

Since the perturbation is decomposed in two temporal contributions, the
ionization amplitude can be written as the sum of the ionization amplitudes
corresponding to each component of the pulse $T_{if}=T_{1}+T_{2}$ and
consequently, the momentum distribution reads: 
\begin{equation}
\frac{\mathrm{d}P}{\mathrm{d}\mathbf{k}}=|T_{1}|^{2}+|T_{2}|^{2}+2%
\,|T_{1}||T_{2}|\cos {\Phi }  
\label{P1}
\end{equation}%
where each term $T_{i}$ (with $i=1,2$) can be expressed as in Eq. (\ref{TSM}%
). Then, the first two terms correspond to the individual emission
probabilities that have the typical structure of ATI peaks at electron
energies following Eq. (\ref{ein}) with the corresponding ponderomotive
energy values $U_{p1}=\left( F_{1}/2\omega \right) ^{2}$ and 
$U_{p2}=\left(F_{2}/2\omega \right) ^{2}$. The third term accounts for the
interference between the two contributions that we call \emph{interpulse
interference} term. The phase $\Phi $ results
\begin{eqnarray}
\Phi &=&\frac{N_{1}\pi }{\omega }\left( I_{p}+\frac{k^{2}}{2}+U_{p1}\right) +%
\frac{N_{2}\pi }{\omega }\left( I_{p}+\frac{k^{2}}{2}+U_{p2}\right)  \notag
\\
&&+\Delta t\left( I_{p}+\frac{k^{2}}{2}\right) +\frac{k}{\omega ^{2}}%
(F_{1}-F_{2})\,.  \label{phi}
\end{eqnarray}

The emission probability will have maximum and minimum values when $\Phi
=\ell \pi $. According whether $\ell $ is odd or even the interference will
be destructive or constructive, respectively. In these cases the
emission probability is
\begin{equation}
\frac{\mathrm{d}P}{\mathrm{d}\mathbf{k}}=\left( |T_{1}|+(-1)^{\ell
}|T_{2}|\right) ^{2}  \label{prob}
\end{equation}%
This take place at $k_{\ell }$ values, that are solutions of the quadratic
equation $\Phi =\ell \pi $, where $\Phi$ is given by Eq. (\ref{phi}).
Let us consider the separation between two successive \textquotedblleft interpulse
interference-peaks" \textcolor{black}{(even $\ell$)} in the PE spectra. 
It is easy to see that
\begin{equation}
2\pi =(k_{\ell +2}^{2}-k_{\ell }^{2})\,\frac{\pi (N+M)}{2\omega }%
+(k_{\ell +2}-k_{\ell })\frac{(F_{1}-F_{2})}{\omega ^{2}}  \label{deltak}
\end{equation}%
This pattern \textcolor{black}{(a sequence of minima and maxima every time that the electron momentum verifies $\Phi=\ell\pi$)} is superimposed to the sum of individual PE
spectra originating from each pulse contribution. This interpulse
interference pattern is observed as fine structure on each
intercycle peak.

In particular, when we consider two identical steps ($F_{1}=F_{2}$ and $%
N_{1}=N_{2}$), the ionization probability in Eq. (\ref{P1}) reads 
\begin{equation}
\frac{\mathrm{d}P}{\mathrm{d}\mathbf{k}}=4\,|T_{1}|^{2}\,\cos ^{2}{(\Phi /2)}
\label{P2}
\end{equation}%
and the phase is reduced to 
$\Phi  =t_{2}\left( I_{p}+\frac{k^{2}}{2}\right)+t_{1}\,U_{p1}$, or what is the same
$\Phi  =t_{1}\left( I_{p}+\frac{k^{2}}{2}+U_{p1}\right)+\Delta t\left(
I_{p}+\frac{k^{2}}{2}\right)$.
In this situation, the interpulse interference is clearly
marked as an oscillating modulation over the single PE spectrum
 and it becomes destructive or constructive according to the time
delay between cycles. 



\subsection{\label{sec:level3}Analysis of PE spectra produced by a few-step envelope pulse}


In this section we evaluate the PE spectra for ionization of the hydrogen
atom from the $1s$ state by laser pulses with envelope defined in Eq. (\ref{f2}%
), using the SCM approximation described above. This will give evidence of
the interferences between cycles with different ponderomotive energy.

\begin{figure}[tbp]
\includegraphics[width=0.5\textwidth]{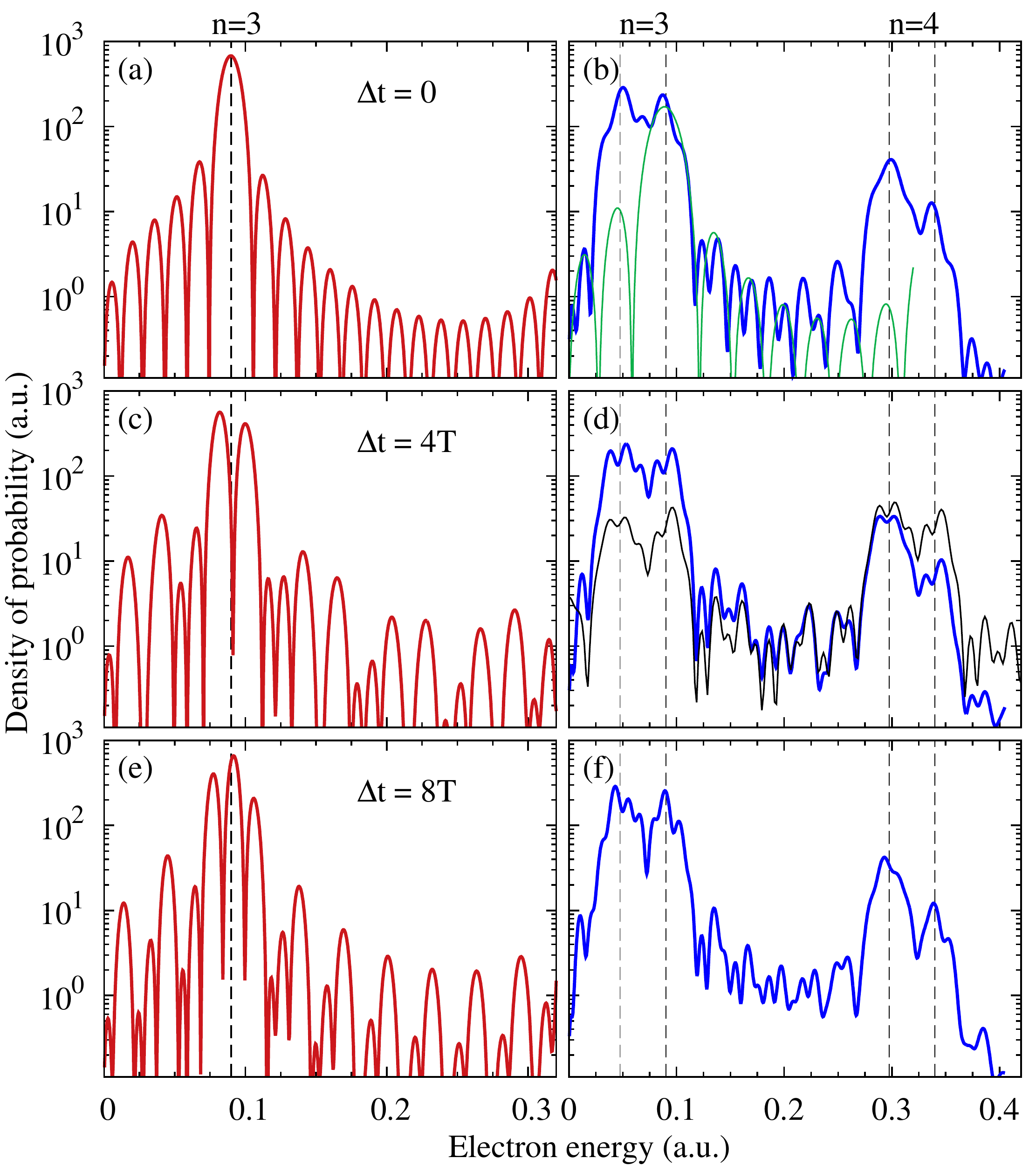}
\caption{(Color online) H(1s) ATI PE spectra obtained within the SCM by laser
pulses with envelope given by Eq. (\protect\ref{f2}) with $\protect\omega=0.25$
a.u., $N_1=N_2=8$ cycles and $F_1=0.2$ a.u.. Left column (red lines): two
identical pulses case with $F_1=F_2$. Right column (blue lines): 
$F_2=0.225$ a.u.. Vertical dashed lines
indicates the position of the ATI peaks following Eq. (\protect\ref{ein})
with $U_{p1}$ and $U_{p2}$ values. The upper labels $n=3$ or $4$ indicate
the number of absorbed photons. In (a-b), (c-d) and (e-f) cases the time
delay between the two parts of the pulse are $0, 4$ and $8$ cycles, respectively.
In (b) PE spectrum for a flat top with 8 cycle pulse in thin (green) line.
In (d) CVA PE spectrum with thin (black) line. }
\label{fig:1}
\end{figure}

In Fig. \ref{fig:1} we present the PE spectra for $F_{1}=0.2$ a.u. and 
$F_{2}=0.2$ a.u. and $0.225$ a.u., in left  and right column
respectively and variable time delay $\Delta t$,
while keeping the frequency constant $\omega =0.25$ a.u. and the number of cycles 
$N_{1}=N_{2}=8$. The rows of Fig. \ref{fig:1} correspond to different temporal
delays between the two pulses. In the first row we show the PE spectra
for the not delay ($t_{2}=t_{1}$) case. 
In (a), as there is no time delay, i.e., $\Delta t = 0$
the two half pulses with amplitude $F_1=F_2=F_0$ merge in a flat top pulse [Fig. \ref{Vs_env} (a)]
and the ponderomotive energy can be uniquely defined as $U_p=U_p^0=(F_0/2\omega)^2$
[see Fig. \ref{Vs_env} (b)] and the PE spectrum is equivalent 
to that obtained from a $N=16$ cycle laser pulse and 
the ATI peak observed corresponds to the absorption of $n=3$ photons 
in agreement with Eq. (\ref{ein}).
In turn,  when the two half pulses have different amplitudes [Fig. \ref{fig:1} (b)],
we observe the presence of ATI peaks at electron energies (marked with vertical
dashed lines) according to Eq. (\ref{ein}) with $U_{p1}=\left( F_{1}/2\omega
\right) ^{2}$ and $U_{p2}=\left( F_{2}/2\omega \right) ^{2}$ values.
The energy separation between the two peaks agrees with the prediction 
$\Delta E = U_{p2}-U_{p1}=(F_2^2-F_1^2)/4\omega ^2 = 0.0425$ a.u..
In Fig. \ref{fig:1} (b) we add with a thin (green) line the PE spectrum for the
smaller pulse alone ($F_{1}=0.2$ a.u. and $F_{2}=0$).
We observe that the presence of the field with
amplitude $F_{2}$ not only produces an ATI peak at $E=0.09$ a.u, according
to Eq. (\ref{ein}) with $U_{p2}=\left( F_{2}/2\omega \right) ^{2}$, but also
the interference structures at the main and lateral peaks in agreement
with the theory of interpulse interference described in the last subsection.

When the interpulse delay increases (middle and lower rows) we observe a
subtle oscillating structure due to the interpulse interference term: the
energy separation between consecutive peaks decreases as increasing $M$ 
[Eq. (\ref{deltak})]. This oscillating pattern is maximum when amplitudes are
equals (left column) and diminishes as the amplitude difference 
$|F_{2}-F_{1}|$ increases (right column).
In the former the interference leads to a null
emission probability at certain energy values following Eq. (\ref{P2}).

In order to check the validity of the employed model 
we compare the SCM and the CVA PE spectra in Fig \ref{fig:1} (d).
We observe that despite SCM overestimates the emission probability,
both calculations results in practically the same interference pattern.

Summarizing, from this simple example  we can distinguish
two different effects on the PE spectrum due to the time-dependent envelope:
(i) different ATI peaks are expected at different energy positions
depending on the field strengths, or what is the same, the emission time intervals,
and (ii) 
\textcolor{black}{ different field strengths and}
the temporal delay 
provoke an interference pattern that is maximal
when the pulses are identical affecting the shapes of the ATI peaks.
The first conclusion (i) can be easily deduced  from the individual contributions 
$|T_i|^2$ in Eq. (\ref{P1}), whereas the second (ii) has its origin
in the interpulse interference term.

In the same way as the simple cases mentioned above,
the ionization amplitudes for pulses with a
several-step envelope could also be written as sum of contribution from each
part of the pulse, and this leads to interferences in the emission
probability. In this situation the number of interfering terms will be
large, the associated peaks superpose and a possible recognition could be
lost. However, the general analysis for the two step pulse can be easily extrapolated to
several-step pulses and therefore, a more deep understanding of the shape of
the spectra could be achieved.

\begin{figure}[tbp]
\includegraphics[width=0.5\textwidth]{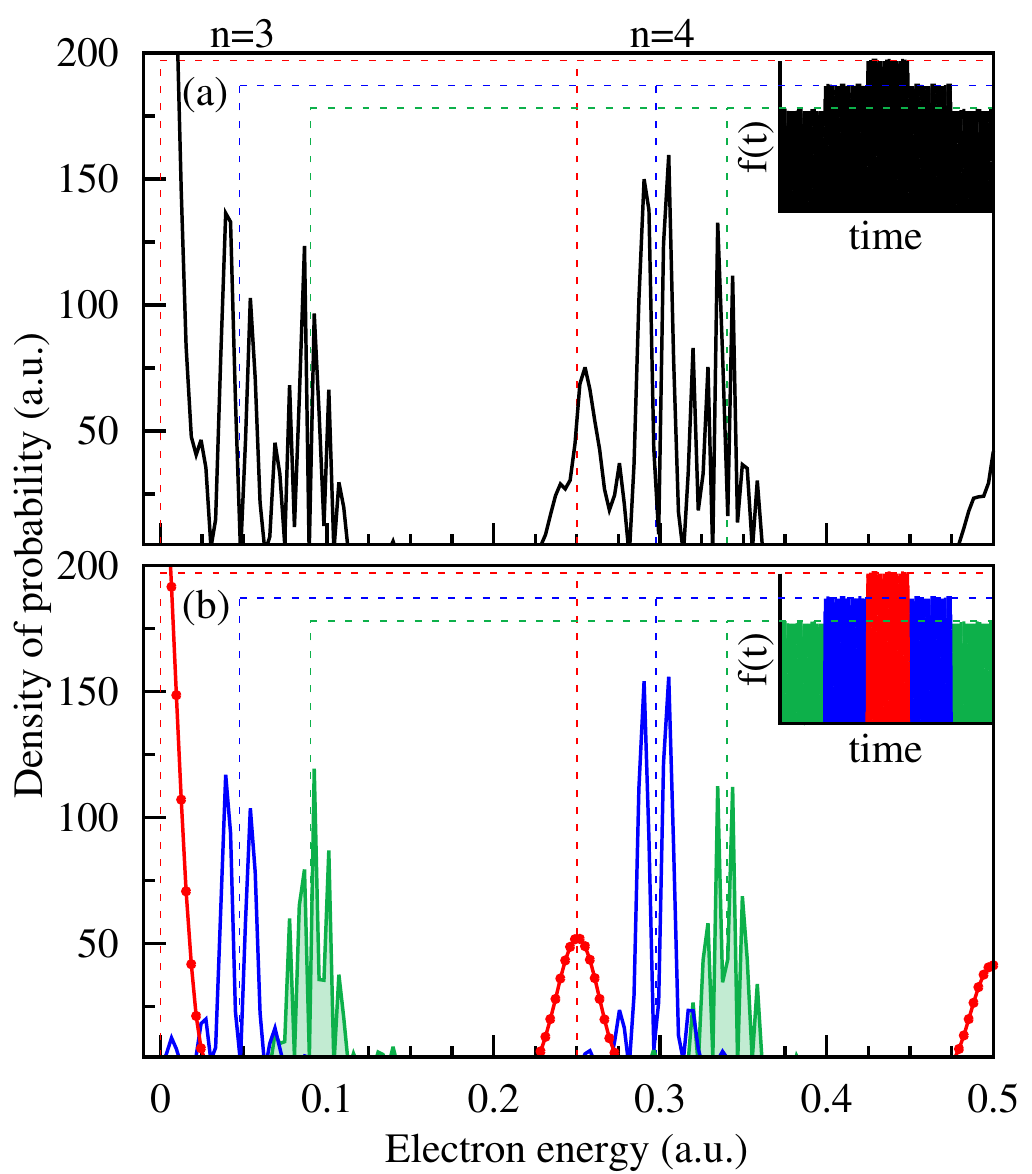}
\caption{(Color online) CVA H(1s) ATI PE spectra with $F_{0} =\protect\omega%
=0.25$ a.u. and envelope shown in the inset. (a) CVA PE spectrum for 5-level
envelope. (b) PE spectra for one step of amplitude $F_0$ with (red) line
with dots, for two step of amplitude $0.9 F_0$ delayed 8 cycles with (blue)
thick line and for two step of amplitude $0.8 F_0$ delayed 24 cycles with
(green) filled curve line. }
\label{fig:2a}
\end{figure}

In the Fig. \ref{fig:2a} we show the PE spectra corresponding to a
symmetric pulse with three different field amplitudes
$0.8F_{0}$, $0.9F_{0}$ and $F_{0}$, as shown in
the inset of figure \ref{fig:2a} (a).
Each step involves 8 cycles, $F_{0}=0.25$ a.u. and $\omega=0.25$ a.u..
As before, we can consider that this pulse
is a superposition of several short pulses with different amplitudes.
Therefore, the resulting electron emission probability results from 
the sum of 5 terms $|Ti|^2$ and 10 interference terms 
($T_{i}T_{j}^{\ast }+T_{i}^{\ast }T_{j}$) with $1\leq i,j\leq 5$ and $i\neq j$.
As before, the main contribution to the
interference effects will come from the amplitudes corresponding to pair of
steps with the same electric field amplitude, temporarily delayed. We
evaluate independently each of these pair of steps and the results are displayed
in Fig \ref{fig:2a} (b). There we observe that the position of the ATI peaks are
given by Eq. (\ref{ein}) with their respective ponderomotive energies, 
as it is indicated by the broken lines.
A comparison of Fig \ref{fig:2a} (a) with \ref{fig:2a} (b) clearly shows that
the contributions from the time delayed interference terms between $T_{1}$
and $T_{5}$ and between $T_{2}$ and $T_{4}$ dominate the respective ATI
peaks. These interferences have an analytical representation as given by
Eq. 
(\ref{P2}).
Furthermore, we have checked that the ATI peak at $0.25$ a.u.
due to the central highest step (red contribution in inset of Fig \ref{fig:2a}(b))
reproduces the non-symmetric shape that it acquires in Fig. \ref{fig:2a}(a) when 
the neighboring level of lower amplitude (blue contribution in inset of Fig \ref{fig:2a}(b)) is also considered (not shown).
An important consequence of  conclusion (i) is that we can consider 
that ATI peaks have a maximum spread directly related to the spread of
the ponderomotive energy values. In the case of Fig \ref{fig:2a} 
this means that the ATI peak for $n=4$ photon absorption 
is comprised in the electron energy range between $0.25$ to $0.34$ a.u.

\subsection{The PE spectra for a continuous envelope}


We consider an electric field with a more realistic envelope described
by a continuous smooth envelope [Eq. (\ref{f3})].
Following the SCM, electron trajectories released at different ionization
times interfere producing an interference pattern. These
ionization times can be easily calculated as $\mathbf{k}+\mathbf{A}(t_{r})=0$.
Unlike the flat-top pulse, ionization times are not equally spaced in the 
continuous envelope case, i.e., sine-squared envelope (see Fig. \ref{Vs_env}(e)).
The temporal separation of these slits depends on the
electron energy and the particular envelope function.
Only the cycles with sufficient intensity will significantly contribute
to the electron emission with a given momenta $\mathbf{k}$. In other
words, unlike Eq. (\ref{deltaS}) for a flat top pulses, the accumulated
action $\Delta S$ does depend on each particular cycle. The action
in Eq. (\ref{action}) depends on the shape of the envelope and the number of cycles
involved, and it does not scale linearly with the cycle order. Therefore, the
intercycle interference will not be described by a simple periodic function
as before and the electron spectra will not exhibit a homogeneous periodic
interference pattern. Anyhow, we are able to analyze qualitatively the PE spectrum
in view of the previous analysis.

\begin{figure}
\includegraphics[width=0.5\textwidth]{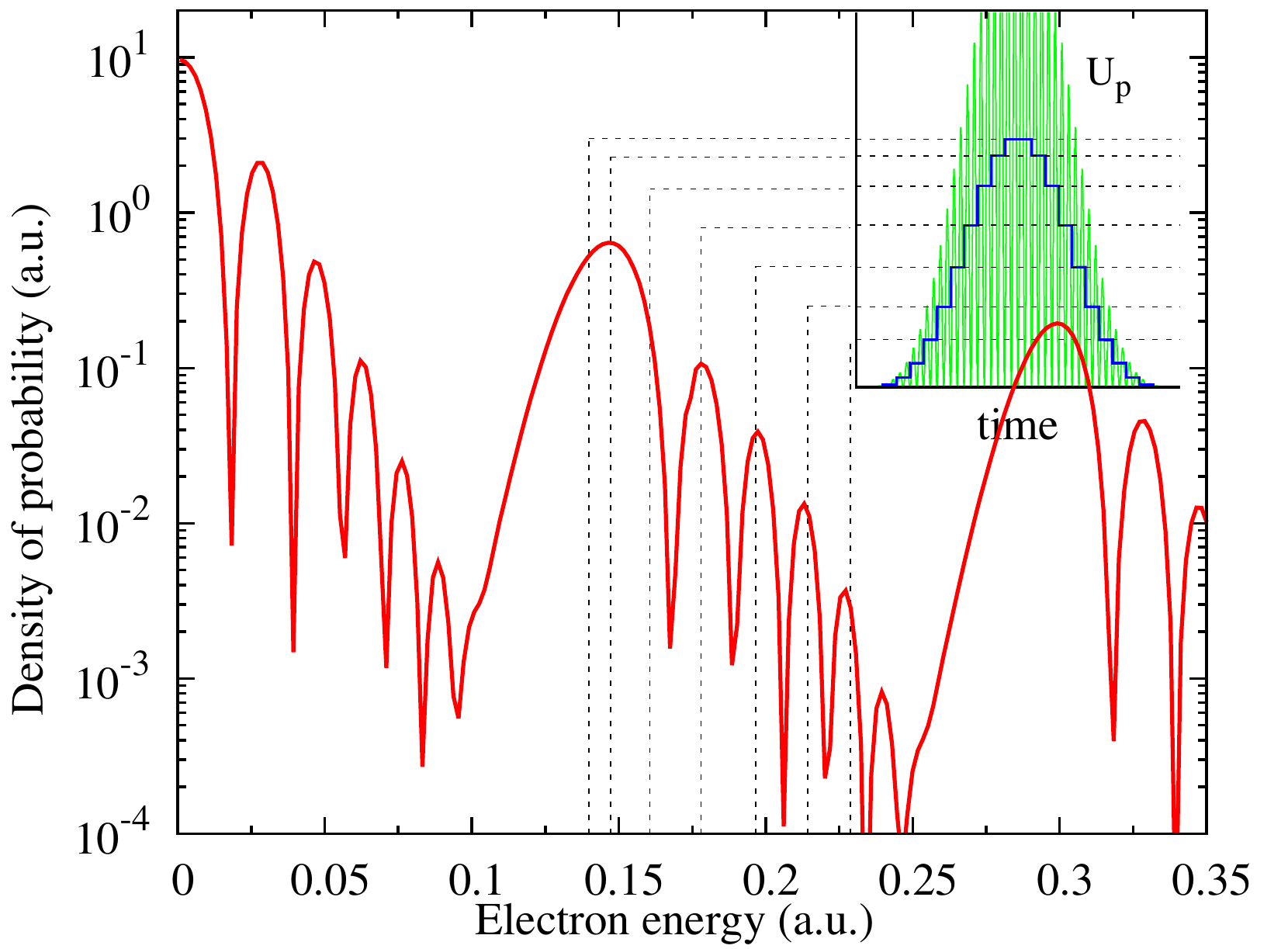} 
\caption{(Color online) H(1s) CVA ATI PE spectra for sinusoidal envelope with 
$F_{0}=0.1$ a.u., $\protect\omega =0.15$ a.u. and $N=24$ cycles.
In the inset we show the quiver (in thin green line) and the cycle dependent ponderomotive energy (in thick blue line) as function of time. }
\label{sine}
\end{figure}

The electron emission will be maximum near the center of the pulse ($t \simeq \tau/2)$,
that is for the strongest values of the field.
Therefore, in the atomic process of absorption of $n$
photons, the kinetic energy will have an lower bound limit given by
Eq. (\ref{ein}), with ponderomotive energy $U_p^0 = (F_0/2\omega)^2$.
The electrons emitted by lower field values in the pulse will feel a lower
ponderomotive energy  and therefore will acquire a higher kinetic energy
with decreasing emission probability. 
In other words, the minimal strength electric field at the beginning and the end
of the pulse produces negligible contribution to the PE spectrum at maximal
value of the electron energy $E_n=n\omega -I_{p}$.
\textcolor{black}{In the middle between both situations, as the field envelope vary continuously the ponderomotive energy is strong dependent on the cycle $(t_0,t_0+2\pi/\omega)$ where the quiver energy average is perform. Depending on the time $t_0$ we can expect continuous values of $U_p$ between $0$ and $U_p^0$ and therefore 
some structures in the PE spectrum are expected in the energy range
$(n\omega -I_p - U_p^0, n\omega -I_p)$ where  effect (ii) described in previous section, is  also added.}
Then, the width of the energy range of
each peak is the peak ponderomotive energy $U_p^0$, provided $U_p^0 < \omega$,
and therefore it is independent of the pulse duration.
As an example, we display in Fig. \ref{sine} the CVA PE spectrum produced by 
a sinusoidal pulse with $\omega =0.15$ a.u., $F_{0}=0.10$ a.u. and $N=24$.
There it can be observed that
the ATI peak corresponding to $n=5$ absorbed photons starts at electron energy 
$5\omega -I_p - U_p^0 = 0.139$ a.u. (when the field is maximum) 
and finishes at $5\omega -I_p = 0.25$ a.u..
In the inset of the figure, \textcolor{black}{as in Fig. \ref{Vs_env} (f),}
we show the cycle averages of the quiver energy,
\textcolor{black}{ \textit{i.e.} the ponderomotive energy,}
 which
are associated to the electron energies indicated by dashed lines in the main
graph. 
\textcolor{black}{In this case, they coincide whit the substructures peaks.}

\begin{figure}
\includegraphics[width=0.5\textwidth]{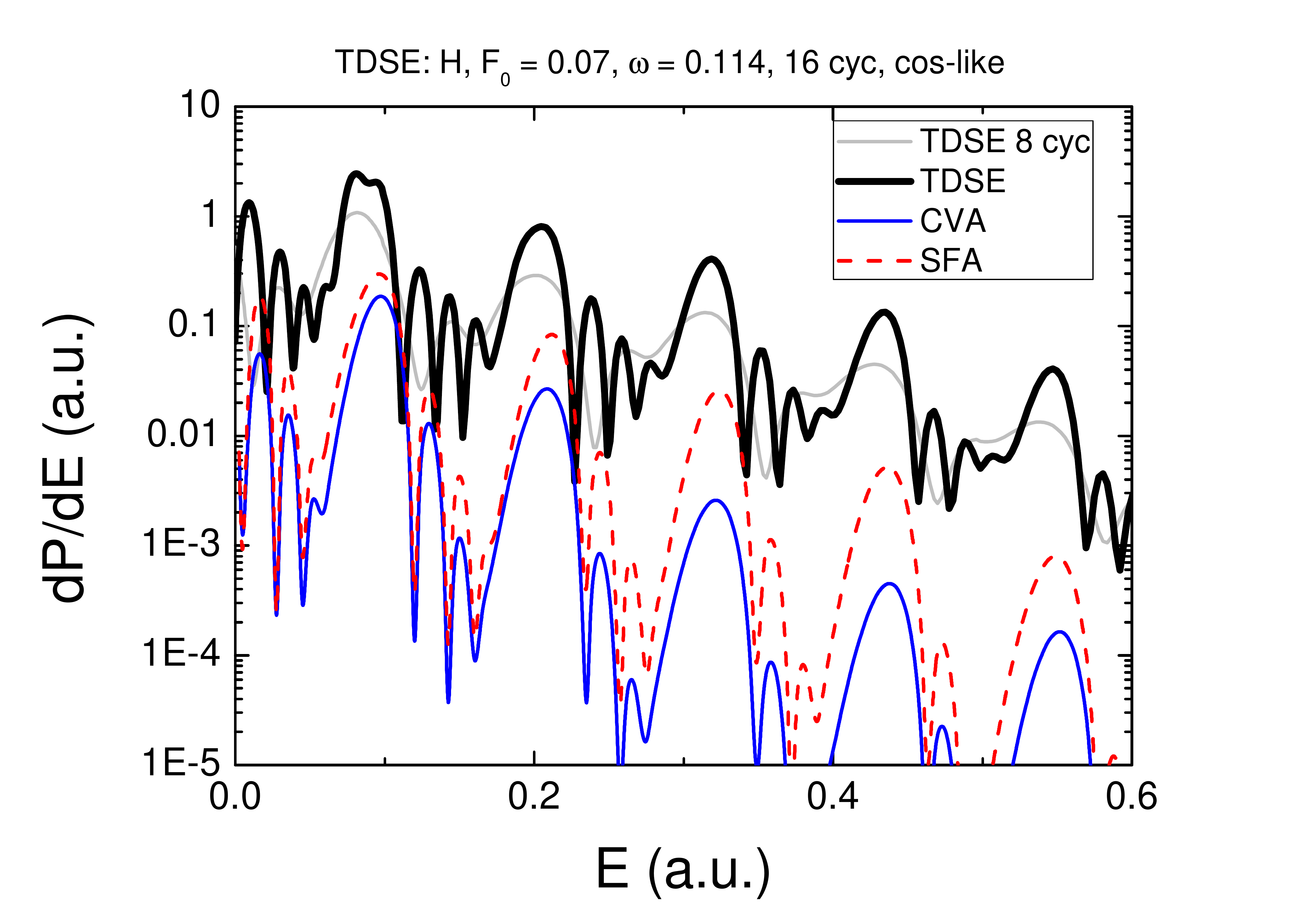} 
\caption{(Color online) H(1s) TDSE (black thick line), CVA (blue thin line),
and SFA (red dashed line) PE spectra for sinusoidal envelope with 
$F_{0}=0.07$ a.u., $\protect\omega =0.114$ a.u. and $N=16$ cycles.
}
\label{TDSECVASFA}
\end{figure}

These results are corroborated with the \textit{ab initio} solution of the
TDSE in the SAE approximation \cite{tong97,tong00,Arbo06prl}. The existence of
the shifted ATI peaks with different values of the ponderomotive energy appears also
in the TDSE calculations (Fig. \ref{TDSECVASFA}). The exact positions of the peaks
do not exactly coincide with the CVA and SFA calculations, 
which happens to be slightly displaced to higher energies. 
The comparison between CVA and SFA PE spectra indicates that the effect
of the long-range Coulomb potential, which is accounted approximately 
in CVA and neglected in SFA, does not have any significant effect in the formation 
of the substructure by a rapidly changing ponderomotive energy \cite{Wickenhauser:06}.


\section{Conclusions}


The quiver energy acquired by an atomic electron
in a laser field is a time dependent quantity, 
\textcolor{black}{that produces a strong time variation of the electron kinetic energy.  
For a simple interpretation of the emitted electron
spectra it is usual to introduce the ponderomotive energy as the time average of the quiver energy,
and use it in a stationary energy conservation rule. 
We show that this approximation is reasonable for pulses with flat envelopes, 
but it is poor for fast changing pulse envelopes.}
Firstly, we consider multiple-step pulses with an envelope that is constant in different 
regions of the time domain and show that each step of constant envelope leads to a 
peak in the PE spectra with position determined by Eq. (\ref{ein}) 
with the corresponding step
ponderomotive energy $U_{p}$. 
We found that the electron emission amplitudes produced by the steps interfere with each other and 
produce a new pattern additional to the inter-cycle peaks. 
We derive an analytic expression for these interferences for the
\textcolor{black}{two} step envelope and discuss how the PE spectrum varies as a function of the
relative field strength and the time delay between \textcolor{black}{steps}. 
When the pulse has a non constant continuous envelope function the intensity 
of the electric field and 
\textcolor{black}{ponderomotive energy}
changes concurrently. 
We found that the electron ionization energy spectra  
\textcolor{black}{is characterized by well known ATI peaks with a width (or substructures)
associated to the variation of ponderomotive energy values. Therefore,}
 the shape of these peaks follows
the shape of the envelope.

\begin{acknowledgments}
Work supported by CONICET PIP0386, \textcolor{black}{PICT-2012-3004 and} PICT-2014-2363 of ANPCyT (Argentina),
the University of Buenos Aires (UBACyT 20020130100617BA).
\end{acknowledgments}

\bibliographystyle{plain}
\bibliography{biblio}
\nocite{*}

\end{document}